\documentstyle[jair,twoside,11pt,theapa,amsmath,amssymb,amsbsy,amsthm,color]{article}
\input epsf  

\newcommand{\set}[1]{\{#1\}}

\jairheading{}{}{}{}{} \ShortHeadings{Demand-Supply Optimization for
a Multi-Connection Reservoir Network} {Chatpatanasiri \& Sriburi}
\firstpageno{1}

\begin{document}

\title{Demand-Supply Optimization with Risk Management for\\ a Multi-Connection Water Reservoir Network}

\author{\name Ratthachat Chatpatanasiri
 \email ratthachat.c@gmail.com \\
       \addr Department of Computer Engineering, Chulalongkorn
       University, Pathumwan, Bangkok, Thailand.
       \name Thavivongse Sriburi \email Thavivongse.S@chula.ac.th \\
       \addr Institute of Environmental Research, Chulalongkorn
       University, Pathumwan, Bangkok, Thailand.}
\maketitle

\begin{abstract}
In this paper, we propose a framework to solve a demand-supply
optimization problem of long-term water resource allocation on a
multi-connection reservoir network which, in two aspects, is
different to the problem considered in previous works. First, while
all previous works consider a problem where each reservoir can
transfer water to only one fixed reservoir, we consider a
multi-connection network being constructed in Thailand in which each
reservoir can transfer water to many reservoirs in one period of
time. Second, a demand-supply plan considered here is static, in
contrast to a dynamic policy considered in previous works. Moreover,
in order to efficiently develop a long-term static plan, a severe
loss (a risk) is taken into account, i.e. a risk occurs if the real
amount of water stored in each reservoir in each time period is less
than what planned by the optimizer. The multi-connection function
and the risk make the problem rather complex such that traditional
stochastic dynamic programming and deterministic/heuristic
approaches are inappropriate. Our framework is based on a novel
convex programming formulation in which stochastic information can
be naturally taken into account and an optimal solution is
guaranteed to be found efficiently. Extensive experimental results
show promising results
of the framework.
\end{abstract}


\section{Introduction}
Water resource allocation in a water reservoir network is an
important problem. In several decades, there has been intensive
research on optimization of a reservoir policy in order to optimally
control amount of water for each reservoir to match its demand in
each time period
\shortcite{Yakowitz:WRR79,Yakowitz:WRR82,Archibald:WRR97,Cai:AWR01,Labadie:JWRPM04,Cervellera:EJOR06,Castelletti:Control07,Wardlaw:JWRPM99,Janejira:Paddy05,Reis:WRM06,Jalali:WRM07,Li:WRM08,Wurbs:JWRPM93,Wurbs:Report05}.
In this paper, we focus on a different but related problem, namely,
a \emph{demand-supply optimization problem of water resource
allocation in a complex reservoir network}.

\begin{figure}[t]
\begin{center}
\vskip -0.35in \setlength{\epsfxsize}{3.0in}
\centerline{\epsfbox{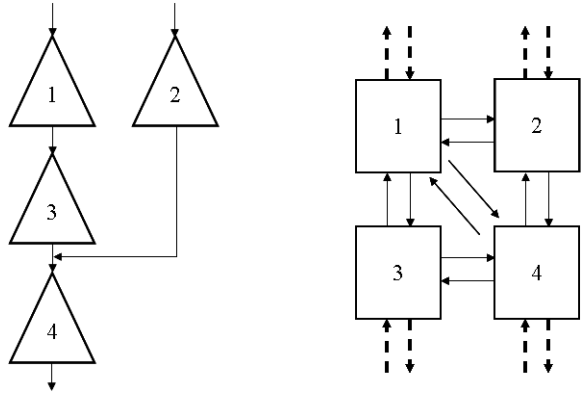}} \vskip -0.1in
\caption{(Left) An example of the traditional reservoir topology
assumed in existing works. (Right) An example of a multi-connection
reservoir topology considered in this work. Solid lines denote
possible transferring directions between two reservoirs. Dashed
lines denote natural inflow and released water for consumers in each
area.} \label{fig_reservoir_topologies}
\end{center}
 \vskip -0.25in
\end{figure}

There are two major differences between ours and a problem of
obtaining a dynamic policy considered in literatures. Firstly,
existing literatures assume a simple network topology of reservoirs
such that (1) a reservoir network is acyclic and (2) water released
from a reservoir directly enters at most one other reservoir. This
assumption is usually implicit in most of previous works but is
stated explicitly by \citeA{Archibald:WRR97}. However, currently a
more complex reservoir network has been being constructed in
Petchburi province, Thailand. In this \emph{multi-connection}
network, each reservoir can transfer water to many reservoirs in one
period of time by using pipes which link between reservoirs.
Figure~\ref{fig_reservoir_topologies} illustrates the difference
between the two topologies. Optimization in a multi-connection
network is dimensionally much higher than a traditional network.
%

Secondly, while the goal of existing works is to obtain a
\emph{dynamic policy of water release}, our goal here is to have
\emph{a static long-term demand-supply plan}. In a problem of
obtaining a dynamic policy, it is assumed that demand is known for
each reservoir in each time period. In our problem, demand (of each
reservoir in each time period) itself will be optimized. Therefore,
our problem can be considered as a prior step needed to be solved
before solving a dynamic policy optimization problem. Moreover, in
our problem, a \emph{risk management} is also taken into account as
there will be a risk (a severe loss) if natural inflow supply cannot
meet the planned supply. More precisely, after achieving a
long-term, e.g. year, demand-supply plan, the obtained plan will be
declared to consumers such as farmers and industries. Consumers then
integrate the declared supply to their year plans. In order to
efficiently utilize their available water, they may invest some
money, e.g. farmers may buy new cattle. Therefore, to prevent a
futile loss of the investment, if real amount of inflow water is
\emph{less} than what is stated in the plan\footnote{Due to the
stochastic nature of the rainfall and run-off phenomena.}, the
government will have to pay an \emph{extra cost}, here called
\emph{a risk}, of making extra water from somewhere outside the
network, e.g. artificial rain, so that the amount of water will be
as specified in the plan.

As a problem is high-dimensional and not dynamic, it is
inappropriate to apply \emph{stochastic dynamic programming} (SDP)
as in previous works
\cite{Cervellera:EJOR06,Castelletti:Control07,Yakowitz:WRR79,Yakowitz:WRR82,Archibald:WRR97,Labadie:JWRPM04,Wurbs:Report05}.
Beside SDP, deterministic and heuristic approaches such as linear
programming, quadratic programming, genetic algorithm and simulated
annealing have also been previously employed
\cite{Cai:AWR01,Wardlaw:JWRPM99,Janejira:Paddy05,Li:WRM08,Jalali:WRM07,Labadie:JWRPM04,Wurbs:Report05}.
Nevertheless, these approaches do not directly support the use of
stochastic information which is fundamental to reservoir network
optimization. In these works, the stochastic information is usually
implicitly employed. For examples, given a probability distribution
of inflow for each reservoir in each time period, the expected value
of inflow can be calculated and input to a deterministic (or
heuristic) method.

In this paper, we propose to apply the framework of \emph{convex
programming} to solve the problem. The nowadays technology of convex
programming is very efficient like least-square and linear
programming, and, with some efforts, many non-linear problems can be
reduced to convex programming formulations  \cite{Boyd:Book04}.
Similar to linear programming, although convex programming is
efficient and appropriate to solve the static problem, a direct
application of convex programming does not naturally support the use
of stochastic information. Here, we present a novel convex
programming formulation which is able to directly take the
stochastic information into account and yield a reasonable output.

The remaining of the paper is structured as follows. In
Section~\ref{sect_prob_spec}, a mathematical formulation of our
problem will be given. In Section~\ref{sect_framework}, related
frameworks and our framework which is proposed to solved the problem
will be explained.  After that, experimental results and discussions
are provided subsequently.




\section{Problem Specification} \label{sect_prob_spec}
Our problem specification here is similar to that of
\citeA{Cervellera:EJOR06}. However, \citeauthor{Cervellera:EJOR06}
considered only a simple network in
Figure~\ref{fig_reservoir_topologies} (left) and did not consider a
static plan with risk management. Let $N$ be the number of
reservoirs in the network and $T$ be the number of periods
considered in the plan, e.g. one year. Let indices $n,m$ run through
$1, ..., N$ and $t$ through $1, ..., T$. At a period of $t$ a
reservoir $n$ can transfer some water to another reservoir $m$ with
amount
\begin{equation} \label{eq_max_transfer}
0 \le q^t_{nm} \le Q_{nm},
\end{equation}
where $Q_{nm}$ is the maximum pumpage capacity in one period between
the two reservoirs. There is also a pumpage cost
\begin{equation} \label{eq_pumpage_cost}
C^t_{nm}(q^t_{nm}),
\end{equation}
where $C^t_{nm}(\cdot)$ is a convex and non-decreasing function.
Aside from inputs from other reservoirs, each reservoir will also
have its natural \emph{inflow} input,
\begin{equation} \label{eq_random_inflow}
x^t_n \sim Pr(x^t_n),
\end{equation}
which is a random variable with respect to a \emph{known}
probability distribution $Pr(x^t_n)$. For each month, there is a
profit from releasing water to consumers
\begin{equation} \label{eq_profit}
G^t_{n}(g^t_{n}),
\end{equation}
where $G^t_{n}(\cdot)$ is a concave and non-decreasing function, and
$g^t_n$ is an amount of released water to consumers. Let $v_n^t$ be
a volume (after transferring $q_{nm}^t$ and releasing $g_n^t$) of
each reservoir on each month. There is a maximum volume $M_n$ for
each reservoir. If $v_n^t < M_n$, according to the traditional rule
of conservation, $v_n^t$ will depend on $g_n^t$, $x_n^t$ and
$q_{nm}^t$. These relations can be concluded by the following state
equation of a volume:
\begin{equation} \label{eq_state}
v_n^t = \min \left\{M_n, v_n^{t-1} - g_n^{t} + x_n^{t} + \sum_{m
\neq n} q^t_{mn} - \sum_{m' \neq n} q^t_{nm'} \right\}.
\end{equation}
And the amount of released and transferred water cannot exceed the
existing volume; hence,
\begin{equation}  \label{eq_release_constraint}
0 \le g^{t}_n + \sum_n q^t_{nm}\le v^{t-1}_n.
\end{equation}
As noted by \citeA{Cervellera:EJOR06},
Eq.~\eqref{eq_release_constraint} is conservative since we do not
include both (uncertain) natural and transferred inflow at period
$t$. In the problem, the initial volume of each reservoir $v^0_n$
has to be specified. Moreover, there are volume constraints at the
end of the optimization period:
\begin{equation}  \label{eq_end_constraint}
v^{T}_n \ge V_n.
\end{equation}
Notice that, unlike previous works, there are two notions of sending
water from a reservoir, i.e. water transfer $q_{nm}^t$ and water
release $g_n^t$. In contrast to water release which is for
consumption, water transfer between two reservoirs requires some
amount of energy for a pumpage. Hence, there is a cost for water
transfer, but a profit for water release.

Finally, we take a risk into account. Since the purpose of a
long-term plan considered in this paper is to optimize the
demand-supply, after obtaining $g_n^t$ from an optimizer, the
resulted $g_n^t$ will be declared to consumers such as farmers and
industries. Hence, $g_n^t$ will be their \emph{target demand} (or, a
\emph{target supply} for the government) since consumers will
integrate $g_n^t$ to their long-term plans. In order to efficiently
utilize their available water, consumers may invest some money, e.g.
farmers may buy new cattle. As a result of a stochastic nature of
$x_n^t$ stated in Eq.\eqref{eq_random_inflow}, \emph{realizable}
amount of release water $\tilde{g}^t_{n}$
may different to its target $g_n^t$.
If $\tilde{g}_n^t < g_n^t$, there will be a severe extra cost, or a
\emph{risk}, of acquiring extra water, e.g. by using artificial
rain, to satisfy the target demand for consumers:
\begin{equation} \label{eq_risk}
R^t_{n}\left( g^t_{n} - \tilde{g}^t_{n} \right)
\end{equation}
where $R^t_n(\cdot)$ is a convex, non-decreasing function and is
zero if $\tilde{g}_n^t \ge g_n^t$.

The goal of this optimization problem is to maximize a profit (or
minimize a loss) subject to all constraints described above where
optimized variables are $\left\{ q^t_{nm}, g^t_n \right\}$.

\section{The Proposed Framework} \label{sect_framework}
In this section, we provide our convex optimization framework to
solve the problem. However, at first, we motivate the necessary of
our framework by illustrating the difficulties of using traditional
methods to solve the problem.

\subsection{The Need of a New Problem Formulation}
From the previous section, it can be observed that our optimization
problem is similar to traditional problem formulations of dynamic
policy optimization; one may temp to think that methods proposed in
previous works can be employed to solve our problem as well. The
purpose of this subsection is to explain reasons that existing
methods are in fact inappropriate for our current problem, and thus
a new method is needed.

We begin by considering the most-popular \emph{stochastic dynamic
programming} (SDP) method
\cite{Yakowitz:WRR79,Yakowitz:WRR82,Archibald:WRR97,Labadie:JWRPM04,Wurbs:Report05,Cervellera:EJOR06,Castelletti:Control07}.
There are some difficulties to solve the problem by using SDP due to
its nature of \emph{dynamic closed-loop control}. Firstly, since a
long-term static plan is our target, all $ \left\{ q^t_{nm}, g^t_n
\right\}_{t=1}^T$ must be obtained at the beginning period $t = 1$.
However, for any $t \ge 2$, a policy obtained from SDP will provide
us optimal values of $q^t_{nm}, g^t_n$ only when the state variables
$\{v^{t-1}_n,v^t_n\}$, which are not known at $t = 1$, are provided
\cite[Chapter 17]{Russell:Book03}. Secondly, the risk defined in
Eq.~\eqref{eq_risk} cannot be straightforwardly taken into account
since, by its definition, a risk occurs because a decision has to be
made \emph{before} knowing the real situation (i.e. a decision is
made before knowing the future values of state variables);
nevertheless, a policy obtained from SDP would provide a decision
only \emph{after} the real situation is known (i.e. a decision will
be made in the future). These difficulties may be crudely solved by
first implementing traditional closed-loop SDP (without the risk) as
illustrated in Figure~\ref{fig_sdp_opt}, and then employ the
\emph{Viterbi} algorithm \cite[Chapter 15]{Russell:Book03} to find
the most-likely sequence of $\{q^t_{nm}, g^t_n\}$. Therefore, a
series of dynamic controls can be converted to a static plan by this
hybrid SDP-Viterbi algorithm. To take the risk into account, some
algorithms may be used to further hybridize the SDP-Viterbi
algorithm. However, this complicated hybrid method of converting
sequential dynamic-controls to a static plan with risk management is
beyond the scope of this paper.

\begin{figure}[t]
\begin{center}
\vskip -0.2in \fbox{
\begin{minipage}{15cm}
\noindent \textbf{Objective}:\\
 $\underset{\{q_{nm}^t,g_n^t\}}{\max} \ \ {\mathbb{E}}_{x} \left[
  \underset{t,n}{\sum} G^t_{n}\left(g^t_{n}\right) -
  \underset{t,n,m}{\sum} C_{nm}^t \left(q_{nm}^t \right) \right]$
                    \\\\
\textbf{subject to}:\\
 $x^t_n \sim Pr(x^t_n)$\\
 $0 \leq q_{nm}^t \leq Q_{nm}$\\
 $0 \le g^{t}_n + \sum_n q^t_{nm}\le v^{t-1}_n$\\
 $v_n^t = \min \left\{M_n, v_n^{t-1} - g_n^{t} + x_n^{t} + \sum_{m \neq n} q^t_{mn} - \sum_{m' \neq n} q^t_{nm'} \right\}.$\\
 $v_n^T \ge V_n$\\
\end{minipage}
} \vskip -0.1in \caption{The traditional SDP formulation of the
multi-connection reservoir optimization problem. This formulation
cannot take a risk into account and is intractable.}
\label{fig_sdp_opt}
\end{center}
\vskip -0.4in
\end{figure}

In fact, we note that there are a third difficulty arising when
using SDP since the SDP formulation illustrated
Figure~\ref{fig_sdp_opt} is not only stochastic and non-linear, but
also high-dimensional. The number of variables grows quadratically
with respect to $N$ (due to the $q_{nm}^t$ variables); thus,
dimensions of the problem is relatively much higher than problems
considered in previous works. Due to the \emph{curse of
dimensionality}
\cite{Bellman:Econo54,Larson:Book68,Bertsekas:Book96} of
discretizing the variables, this high-dimensionality actually
prevents the use of SDP in general; even the most efficient SDP
approximation, the \emph{neuro-dynamic programming} method
\cite{Bertsekas:Book96}, known in literatures can cope with problems
of only 30 dimensions \cite{Cervellera:EJOR06}.

Beside SDP, some existing works proposed to apply deterministic and
heuristic approaches such as linear programming (LP), quadratic
programming (QP), genetic algorithm and simulated annealing for
conventional reservoir networks
\cite{Cai:AWR01,Wardlaw:JWRPM99,Janejira:Paddy05,Li:WRM08,Jalali:WRM07,Labadie:JWRPM04,Wurbs:JWRPM93,Wurbs:Report05}.
Nevertheless, the most important limitation of these methods is that
they cannot directly take the stochastic information and the risk
into account\footnote{Because the risk is defined based on
uncertainty, if uncertainty itself cannot be taken into account, the
risk also cannot be incorporated.}. Usually, uncertainty of inflow
is ignored, and $x_n^t$ is simply replaced by a deterministic value
such as its mean $\bar{x}_n^t$. Henceforth, a deterministic approach
with these simple deterministic value inputs will be called a
\emph{traditional (or conventional) deterministic} method.
Since a heuristic approach does not guarantee an optimal solution,
the traditional deterministic approach is the most suitable method
among existing approaches for solving the problem considered here.

Finally, we note that in the literatures of convex optimization
itself, there is a methodology called \emph{robust programming}
\cite[Chapter 4]{Boyd:Book04} which indeed can efficiently utilize
the stochastic information. Nonetheless, this robust programming is
in fact a worst-case optimization method where the worst situation
will be solved. Specifically, in our reservoir problem, robust
programming will consider the situation where minimum inflow occurs
for each reservoir in each time period. This worst case situation
indeed rarely occurs, and hence robust programming is also
inappropriate for our problem. In the next section, we present a
novel convex optimization formulation which can corporate the
stochastic information in a more-efficient way.

\subsection{Convex Programming Formulation}
Here, we give a new optimization formulation to appropriately solve
the problem stated in Section~\ref{sect_prob_spec} using the convex
programming framework \cite{Boyd:Book04}. Note that in contrast to
stochastic dynamic programs, high-dimensional convex programs which
are deterministic can be solved efficiently. Moreover, the long-term
plan where $ \left\{ q^t_{nm}, g^t_n \right\}_{t=1}^T$ is needed at
the beginning period can be easily achieved.


The most important step here is to take the stochastic inflow
information into this deterministic framework. Our main idea is,
instead of treating stochastic inflow variables as random variables
or constants like previous works, \emph{the stochastic variables
$x_n^t$ themselves will be optimized, as our ``best predictions'' of
inflows}. Then, in order to take the risk into account, we propose
to consider the average risk of Eq.\eqref{eq_risk} over all possible
actual inflow. The best inflow predictions will be optimized so that
they will balance between profits and risks. For instances, if the
risk function $R(\cdot)$ is severe, the best predictions will be
conservative such that they will smaller than the mean, i.e. $x^t_n
< \bar{x}^t_n$; this small-value prediction of $x^t_n$ will limit a
predicted volume $v^t_n$, and hence also limit amount of water
transfer $q^t_{nm}$, i.e., if the risk is costly and we are unsure
of being have abundant of water, we should not transfer the water to
other reservoir carelessly; in contrast, if $R(\cdot)$ return low
values, the best predictions will be aggressive such that they will
be smaller than the mean, i.e. $x^t_n > \bar{x}^t_n$.

To formally explain our framework, we first note that if we are able
to guarantee that for each reservoir and each time period $v^t_{n}
\le M_n$, then the state equation simplifies to
\begin{equation} \label{eq_ideal_state}
{v}^t_{n} = {v}_n^{t-1} - {g}_n^{t} + \tilde{x}_n^{t} + \sum_{m \neq
n} q^t_{mn} - \sum_{m' \neq n} q^t_{nm'}.
\end{equation}
It will be explained below that our problem formulation guarantees
to satisfy the condition $v^t_{n} \le M_n$. Now, suppose an actual
(future observed) inflow is $\tilde{x}_n^t$. Define
$\tilde{g}^t_{n}$ as the realizable amount of water released by a
reservoir $n$ at a period $t$ (see Eq.\eqref{eq_risk})
\begin{equation} \label{eq_real_release}
\tilde{g}^t_{n} = g^t_{n} + \tilde{v}^{t-1}_{n} - v^{t-1}_{n},
\end{equation}
where $\tilde{v}^t_{n}$ is as an actual volume defined recursively
as
\begin{equation} \label{eq_real_state}
\tilde{v}^t_{n} = \tilde{v}_n^{t-1} - \tilde{g}_n^{t} +
\tilde{x}_n^{t} + \sum_{m \neq n} q^t_{mn} - \sum_{m' \neq n}
q^t_{nm'},
\end{equation}
and $\tilde{v}_n^{0} = v_n^{0}$. Eq.\eqref{eq_real_release} states
that the realizable amount of release from the reservoir $n$ at
period $t$ is the target release $g^t_{n}$ plus the difference
between the actual and target volumes $\tilde{v}^{t-1}_{n} -
v^{t-1}_{n}$. By Eq.\eqref{eq_real_release}, we have that
$\tilde{g}^t_{n} - g^t_{n} = \tilde{v}^{t-1}_{n} - v^{t-1}_{n}$ and
$\tilde{v}^{t-1}_{n}- \tilde{g}^t_{n} = v^{t-1}_{n} - g^t_{n}$.
Using these relations and subtract Eq.\eqref{eq_real_state} with
Eq.\eqref{eq_ideal_state}, then, the expected risk
${\mathbb{E}}_{\tilde{x}} \left[ R^t_{n}\left( g^t_{n} -
\tilde{g}^t_{n} \right)\right]$ can be stated as
\begin{equation}
 {\mathbb{E}}_{\tilde{x}} \left[ R^t_{n}\left(  g^t_{n} - \tilde{g}^t_{n} \right)\right] =
 {\mathbb{E}}_{\tilde{x}} \left[ R^t_{n}\left(  x^{t-1}_{n} - \tilde{x}^{t-1}_{n}
 \right)\right].
\end{equation}
For any $t$, to (approximately) calculate ${\mathbb{E}}_{\tilde{x}}
\left[ R^t_{n}\left( x^t_{n} - \tilde{x}^t_{n} \right)\right]$, we
discretize the domain of $Pr(x_n^t)$ into a finite domain set,
${\mathcal{X}}_n^t$:
\begin{equation} \label{eq_approx_risk}
 {\mathbb{E}}_{\tilde{x}} \left[ R^t_{n}\left(  x^t_{n} - \tilde{x}^t_{n} \right)\right] \approx
 \sum_{\tilde{x}_n^t \in {\mathcal{X}}_n^t } Pr(\tilde{x}_n^t) R^t_{n}\left(  x^t_{n} - \tilde{x}^t_{n}
 \right).
\end{equation}
Note that in practice the domain of $Pr(x_n^t)$ is originally
discrete since we are able to collect only finite statistics of
$x_n^t$ and thus Eq.\eqref{eq_approx_risk} can be actually exact.
Note further that since $R^t_n(\cdot)$ is convex,
$\sum_{\tilde{x}_n^t \in {\mathcal{X}}} Pr(\tilde{x}_n^t)
R^t_{n}\left(  x^t_{n} - \tilde{x}^t_{n} \right)$ is also convex.

\begin{figure}[t]
\begin{center}
\vskip -0.2in \fbox{
\begin{minipage}{15cm}
\noindent \textbf{Objective}:\\
 $\underset{\{q_{nm}^t,g_n^t,x_n^t\}}{\max} \ \
  \underset{t,n}{\sum} {G}_{n}^t \left(g_n^t \right) -
  \underset{t,n,m}{\sum} C_{nm}^t \left(q_{nm}^t \right) -
  \underset{t,n}{\sum} F_{n}^t \max\left( v_n^t - M_n  , 0 \right) -
  \underset{t,n}{\sum} {\mathbb{E}}_{\tilde{x}} \left[ R^t_{n}\left(  x^t_{n} - \tilde{x}^t_{n}  \right)\right]$
                    \\\\
\textbf{subject to}:\\
$0 \leq q_{nm}^t \leq Q_{nm}$\\
$0 \le g^{t}_n + \sum_n q^t_{nm}\le v^{t-1}_n$\\
$v_{n}^t = v_{n}^{t-1} - g_{n}^{t}  + x_n^t + \sum_{m \ne n} q_{mn}^{t} - \sum_{m' \ne n} q_{nm'}^{t}$\\
$v_n^T \ge V_n$\\
$\min({\mathcal{X}}_n^t) \le x_{n}^{t} \le \max({\mathcal{X}}_n^t)$\\
\end{minipage}
} \vskip -0.1in \caption{Summary of the proposed optimization
problem.} \label{fig_proposed_opt}
\end{center}
\vskip -0.2in
\end{figure}

Our optimization formulation is summarized in
Figure~\ref{fig_proposed_opt}. This maximization problem formulation
consists of a concave objective function and linear constraints, and
thus the problem can be efficiently solved by the convex programming
framework \cite{Boyd:Book04}. The initial values of $v_{n}^{0}$ has
to be provided.

Note that in order to make our problem formulation convex and
efficient, some constraints in Section~\ref{sect_prob_spec} are
changed and new constraints are added. First, the constraint
$``\min({\mathcal{X}}_n^t) \le x_{n}^{t} \le
\max({\mathcal{X}}_n^t)"$ is added so that $x_{n}^{t}$ will take
only realizable values. Next, the constraint of Eq.\eqref{eq_state}
is simplified by introducing a new cost term ``$F_n^t \max ( v_n^t -
M_n  , 0)$''. Here, $F_{n}^t$ is a big constant, and by setting
$F_n^t$ large enough, this will guarantee that $v_n^t \le M_n$. Note
that to obtain a standard objective function form, this term can be
easily replaced by a  slack variable.

Table~\ref{table_compare} summarizes the differences between our
framework and the other frameworks. In the table, we do not consider
heuristic methods such as genetic algorithm or simulated annealing.
The features of heuristic methods are similar to those of
deterministic methods: the major difference is that heuristic
methods can handle non-linear objective functions. Nevertheless, an
ability to handle a non-linear function comes up with a certain
tradeoff: heuristic approaches are either sub-optimal or
intractable. Also heuristic methods usually were tested with
problems with convex objective functions
\cite{Wardlaw:JWRPM99,Janejira:Paddy05,Jalali:WRM07,Li:WRM08}.
Therefore, our framework presented here can substitute these works
with the main advantage that our framework guarantees to find an
optimal solution with a polynomial running time. We note also that
our framework in fact generalizes traditional uses of LP and QP: if
we specify the $Pr(x_n^{t}  = x') = 1$ for some $x'$ (such as
$\bar{x}_n^{t}$), then by the constraint $\min({\mathcal{X}}_n^{t})
<= x_n^{t} <= \max({\mathcal{X}}_n^{t})$, the best prediction will
be $x'$, and the risk term $R(x_n^{t} - \tilde{x}_n^{t})$ will
become a constant and can be removed from the optimization problem;
for example, in cases that $G(\cdot)$ and $C(\cdot)$ are
linearizable, the resulted program is simply an instance of
traditional LP applied to a multi-connection reservoir network, see
e.g. \citeA{Wurbs:Report05}.

\begin{table}[t] \vskip -0.0in\caption{Summary of differences between the three frameworks.} \label{table_compare}
\begin{center}
\begin{footnotesize}
\vskip -0.25in
\begin{sc}
\vskip -0.05in
\begin{tabular}{p{2.2cm}|p{4.3cm}|p{3.9cm}|p{3.9cm}}
\hline
\   &  Traditional SDP & Traditional LP/QP   &  Our Convex Program \\
\hline
Type          & Closed-Loop Control &  Open-Loop Optimization &  Open-Loop Optimization\\
Policy         & \footnotesize{Dynamic: Real-Time Decision} &  Static: Long-Term Plan &  Static: Long-Term Plan\\
Running Time  & {Intractable} &  Tractable&  Tractable\\
Stochastic   & Included & {Not Included} & Included\\
Objective   & Non-linear & {Linear/Quadratic} & Concave (Maximization)\\
Risk & Not Included & {Not Included} & Included\\
Solution & {Sub-Optimal} & Optimal & Optimal\\
$x_n^t$       & Random Variables & Constants & Optimized Variables\\
$g_n^t$       & Control Variables & Optimized Variables & Optimized Variables\\
$q_{nm}^t$       & Control Variables & Optimized Variables & Optimized Variables\\
\hline
\hline
\end{tabular}
\end{sc}
\end{footnotesize}
\end{center}
\vskip -0.2in
\end{table}

\section{Experiments} \label{sect_exp}


As demonstrated in the previous section (see also
Table~\ref{table_compare}), a deterministic method is the most
appropriate candidate among existing methods which can be
implemented to solve our problem; hence, in this section, we
evaluate our method and compare it to a standard deterministic
method. \textsc{matlab} with the optimization interfaces from
\textsc{Yalmip} \cite{YALMIP} is used for all implementations. For
each simulated situation, the following process
is employed:\\

1. All functions, $C_n^t(\cdot)$, $G_n^t(\cdot)$ and $R_n^t(\cdot)$,
all constants, $Q_{nm}$ and $M_n$, and all probability distributions
$Pr(x_n^t)$ are input to our method (or a deterministic method). The
long-term demand-supply plan of $\set{q_{nm}^t,g_{n}^t}$ is then
obtained for each method.\\

2. A sequence of inflow is generated  $\{\tilde{x}^t_n\}$ where each
$\tilde{x}^t_n \sim Pr(x^t_n)$. $\{\tilde{g}^t_n\}$ is
then calculated according to Eq.\eqref{eq_real_release} and Eq.\eqref{eq_real_state}.\\

3. The total profit of each method will calculated from:
  \[
  \underset{n}{\sum} G^t_{n}\left(g^t_{n}\right) -
  \underset{n,m}{\sum} C_{nm}^t \left(q_{nm}^t \right) -
  \underset{n}{\sum}  R^t_{n}\left(  \tilde{g}^t_{n}  -
  g^t_{n}\right).
   \]

4. Steps 2. and 3. are repeated 100 times to calculate the average
performance of each algorithm\footnote{Note that Step 1. is
deterministic and not depend on a real sequence of inflow
$\{\tilde{x}^t_n\}$.}.

\subsection{Simple Network}
\label{sect_exp1}
\begin{figure}[t]
\begin{center}
\vskip -0.25in \setlength{\epsfxsize}{5.0in}
\centerline{\epsfbox{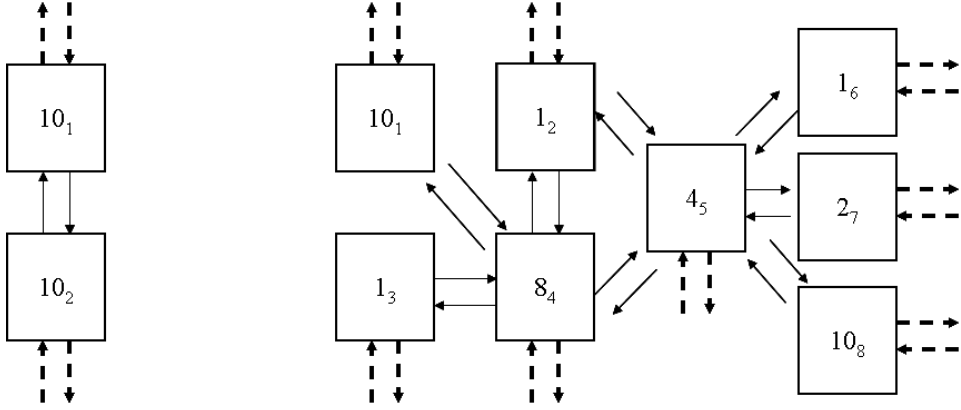}} \vskip -0.1in
\caption{(Left) A simple 2-reservoir network. (Right) The Ang-Puang
multi-connection reservoir network being constructed in Petchburi
province, Thailand. The maximum volume of each reservoir is
indicated. Subscripts denote orders of reservoirs.}
\label{fig_reservoir_exp}
\end{center}
 \vskip -0.25in
\end{figure}

\begin{figure}[t]
\begin{center}
\vskip -0.05in \setlength{\epsfxsize}{7.5in}
\centerline{\epsfbox{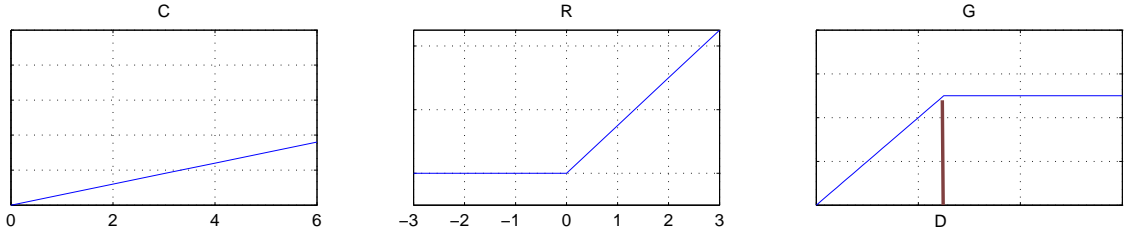}} \vskip -0.1in \caption{The
$C_n^t(\cdot)$, $R_n^t(\cdot)$ and $G_n^t(\cdot)$ employed in the
Section~\ref{sect_exp1} and Section~\ref{sect_exp2}.}
\label{fig_CRG}
\end{center}
 \vskip -0.25in
\end{figure}

\begin{figure}[t]
\begin{center}
\vskip -0.35in \setlength{\epsfxsize}{7.0in}
\centerline{\epsfbox{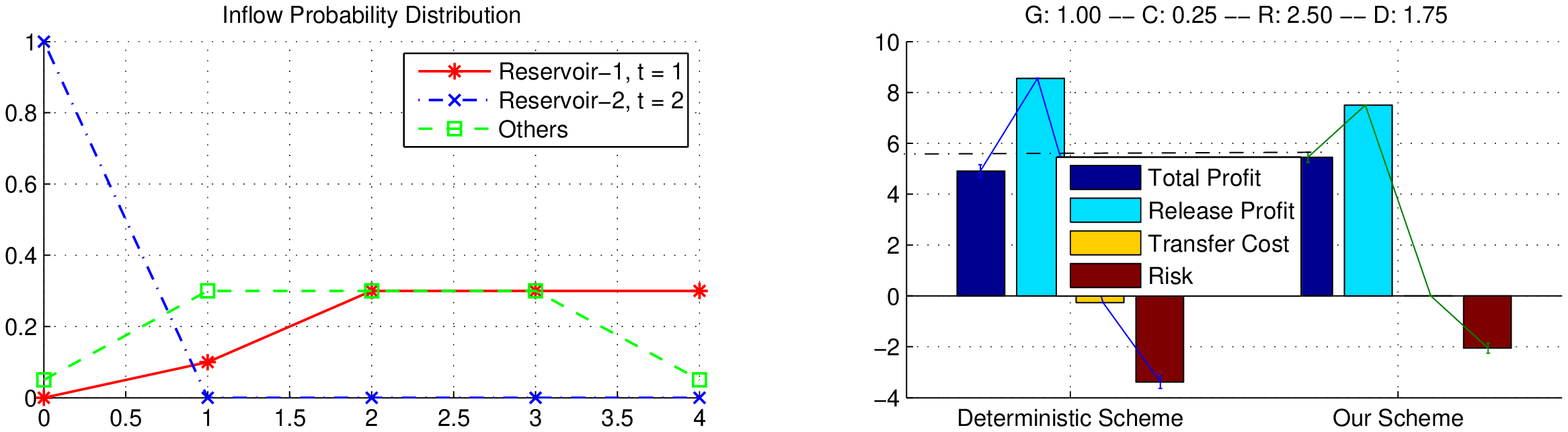}}
\setlength{\epsfxsize}{7.0in}
\centerline{\epsfbox{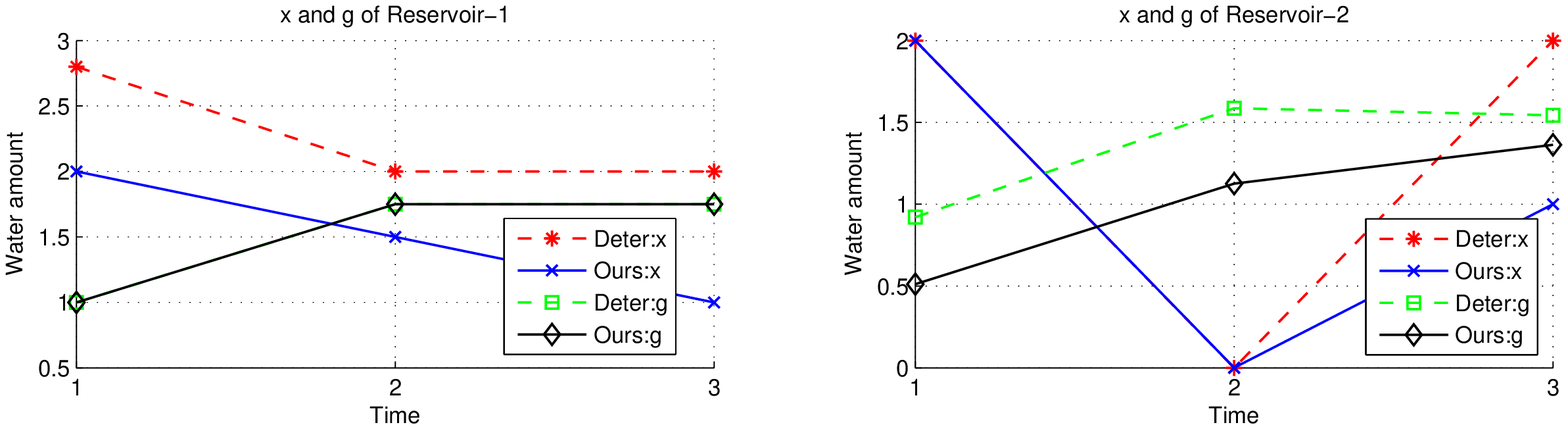}}
 \setlength{\epsfxsize}{7.0in}
\centerline{\epsfbox{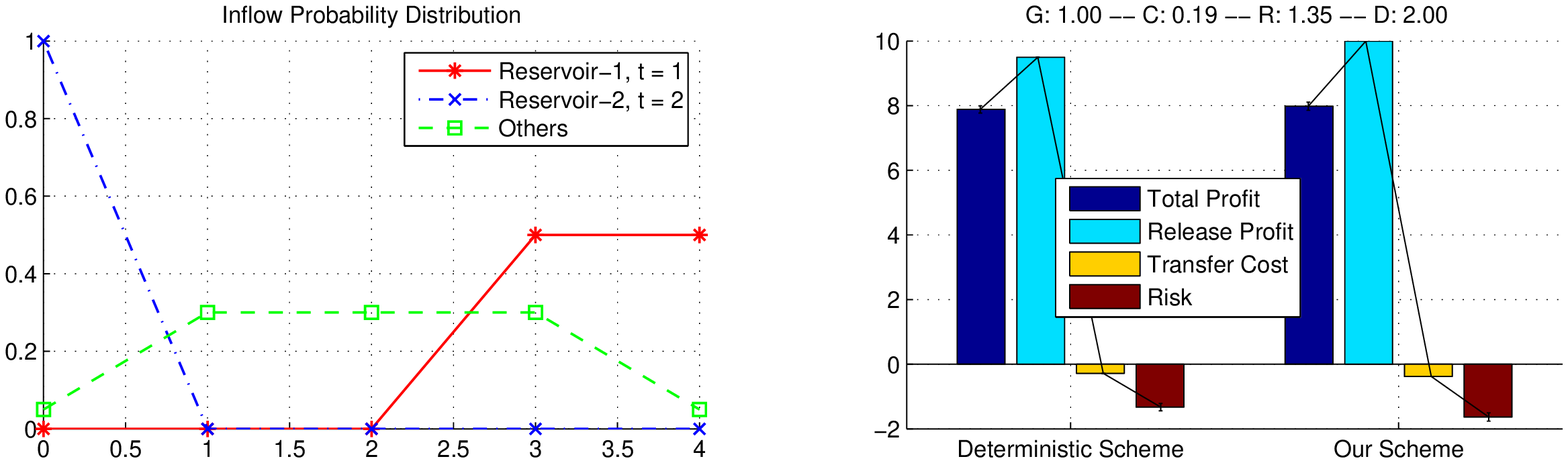}}
\setlength{\epsfxsize}{7.0in}
\centerline{\epsfbox{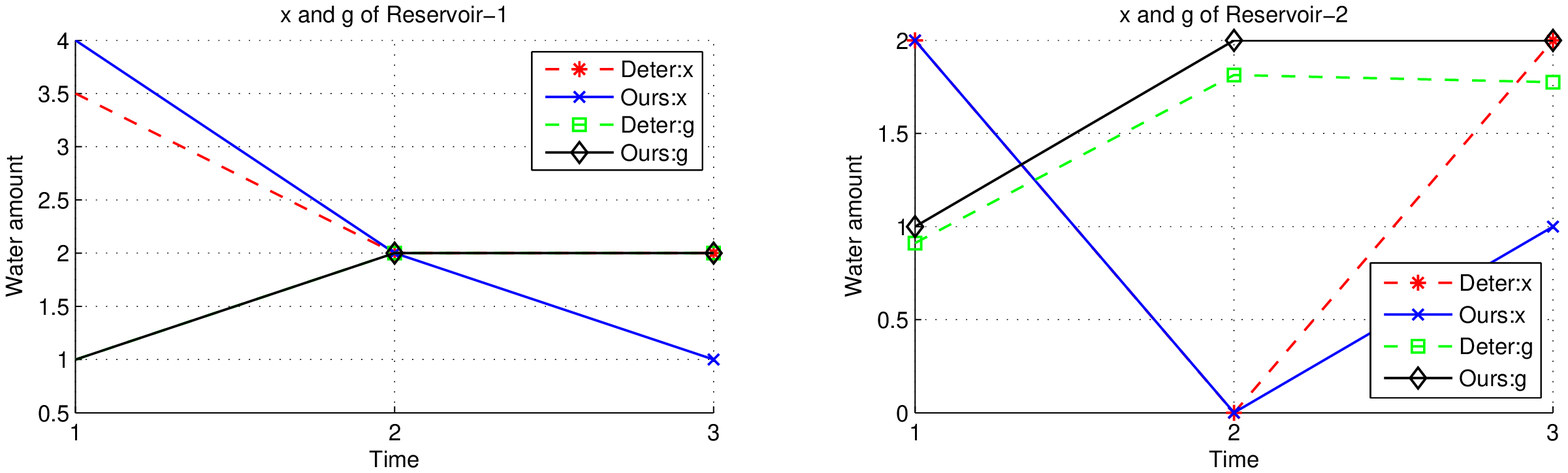}} \vskip -0.1in \caption{[The
Top 4 Figures]: (Up-Left) The probability distributions of inflow
used in the simple network. With the exception of $Pr(x_1^1)$ and
$Pr(x_2^2)$, the distributions are the same for all $n$ and $t$.
(Up-Right) Performance summary: ``(Average) Total Profit = Release
Profit - Transfer Cost - (Average) Risk'' with standard deviations
of ``Total Profit'' and ``Risk''. The C,R,G slopes and the D value
of Figure~\ref{fig_CRG} are also shown on the top. (Down)
$\set{x^t_n}$ and $\set{g^t_n}$ are shown for each reservoir. [The
Bottom 4 Figures]: Another case of the the simple network.}
\label{fig_simple1}
\end{center}
 \vskip -0.25in
\end{figure}

To intuitively understand results obtained from each algorithm, it
is best to start from a a simple reservoir network as illustrated in
Figure~\ref{fig_reservoir_exp} (Left). Here, $C_n^t(\cdot)$,
$R_n^t(\cdot)$ and $G_n^t(\cdot)$ are illustrated in
Figure~\ref{fig_CRG} where several values of their slopes and `$D$'
(shown in the figure of $G_n^t(\cdot)$) will be tested. `$D$' can be
interpreted as the ``maximum demand'' for each reservoir at a
certain period. In this experiment, these functions are the same for
all $n$ and $t$. Since these functions are linearizable, a
deterministic method considered in this example is LP. The following
values are specified: $T = 3$, $N =2$, $M_1 = M_2 = 10$, $Q_{12} =
Q_{21} = 5$, and $v^0_1 = v^0_2 = V_1 = V_2 = 1$.
Figure~\ref{fig_simple1} illustrates two cases of this simple
network and the corresponding results obtained from traditional LP
and our method. The traditional LP implementation can be achieved by
simply replacing each uncertain $x_n^t$ with a deterministic value;
$\bar{x}_n^t$ is used here.

The two cases here illustrate the same simple situation which can be
intuitively explained as follows. The first reservoir has a good
probability to have more than usual inflow at $t = 1$, and the
second reservoir will surely do not have any inflow at $t = 2$.
Therefore, clearly, the first reservoir should send water to the
second reservoir on the second period, but the question is:
\emph{how much} should be transferred? The results obtained from
traditional and our methods can be used as answers to this question.
From the simulations of these two cases (and other several cases not
shown here due to the space limitation), our method always
significantly outperforms traditional LP. The main reason is because
traditional LP \emph{inefficiently} takes stochastic information
into account.

The top 4 figures of Figure~\ref{fig_simple1} illustrate a case
where the risk cost is high, and our method becomes more
conservative about its predictions of inflow. Even though our method
cannot make a high average released profit as that of traditional
LP, our method does provide a much lower average risk cost, and
therefore results in a higher average total profit (5.49) than that
of traditional LP (4.91). The bottom 4 figures of
Figure~\ref{fig_simple1} illustrate a second case where a risk plus
a transferred cost is not severe and, with high probability, $x_1^1$
will be very high. In this case, our method results in aggressive
inflow prediction of $x^1_1$ which in turn leads to a higher release
profit and a higher average total profit (7.99) compared to that of
traditional LP (7.88).

\subsection{Complex Network}
\label{sect_exp2}
\begin{figure}[t]
\begin{center}
\vskip -0.25in \setlength{\epsfxsize}{5.0in}
\centerline{\epsfbox{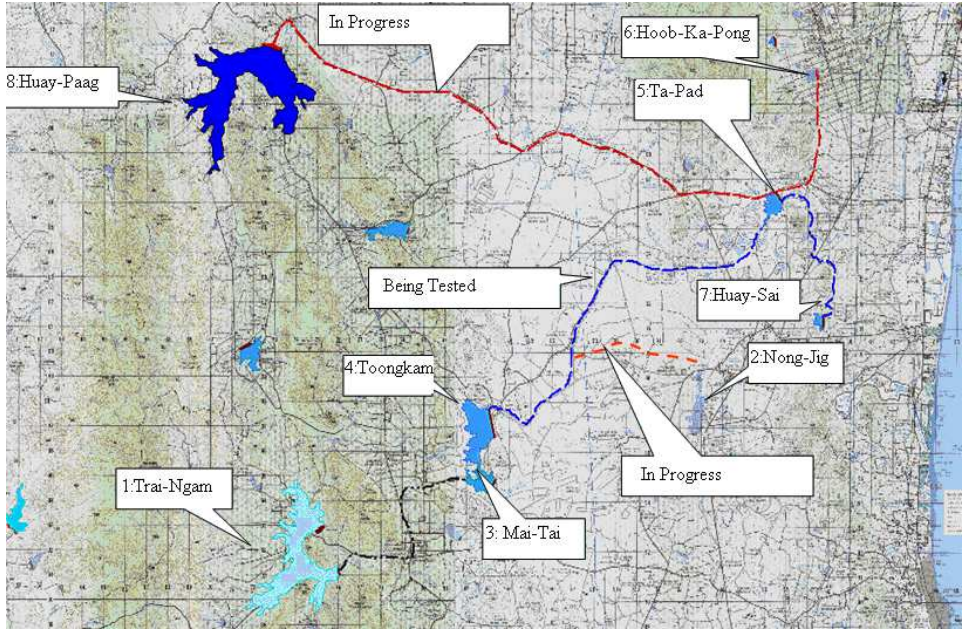}} \vskip -0.1in \caption{The
geography of the Ang-Puang multi-connection reservoir network. ``In
Progress'' denotes a pipe which is being constructed or is planned
to be constructed.} \label{fig_angpuang}
\end{center}
 \vskip -0.25in
\end{figure}

\begin{figure}[t]
\begin{center}
\vskip -0.35in \setlength{\epsfxsize}{7.0in}
\centerline{\epsfbox{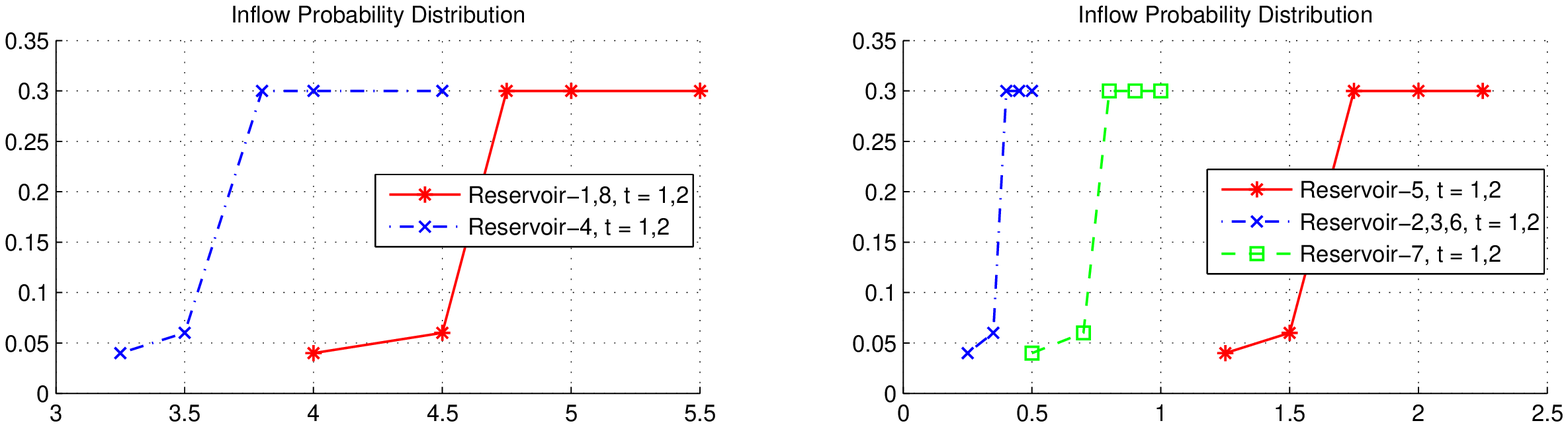}}
\setlength{\epsfxsize}{7.0in}
\centerline{\epsfbox{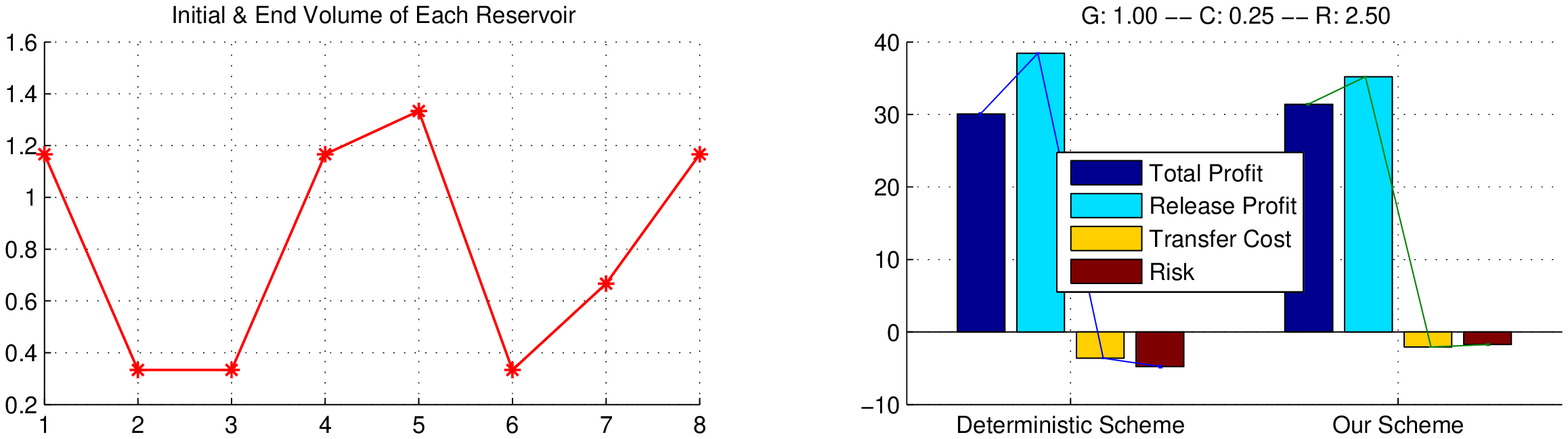}} \vskip -0.1in
\caption{(Up-Left) Inflow probability distributions of the three big
reservoirs for $t=1,2$. (Up-Right) Inflow probability distributions
of the other reservoirs for $t=1,2$. (Down-Left) $v_0$ and $V_n$ for
each reservoir. (Down-Right) Summary of the profit for each method.}
\label{fig_complex1}
\end{center}
 \vskip -0.25in
\end{figure}

\begin{figure}[t]
\begin{center}
\vskip -0.5in \setlength{\epsfxsize}{7.0in}
\centerline{\epsfbox{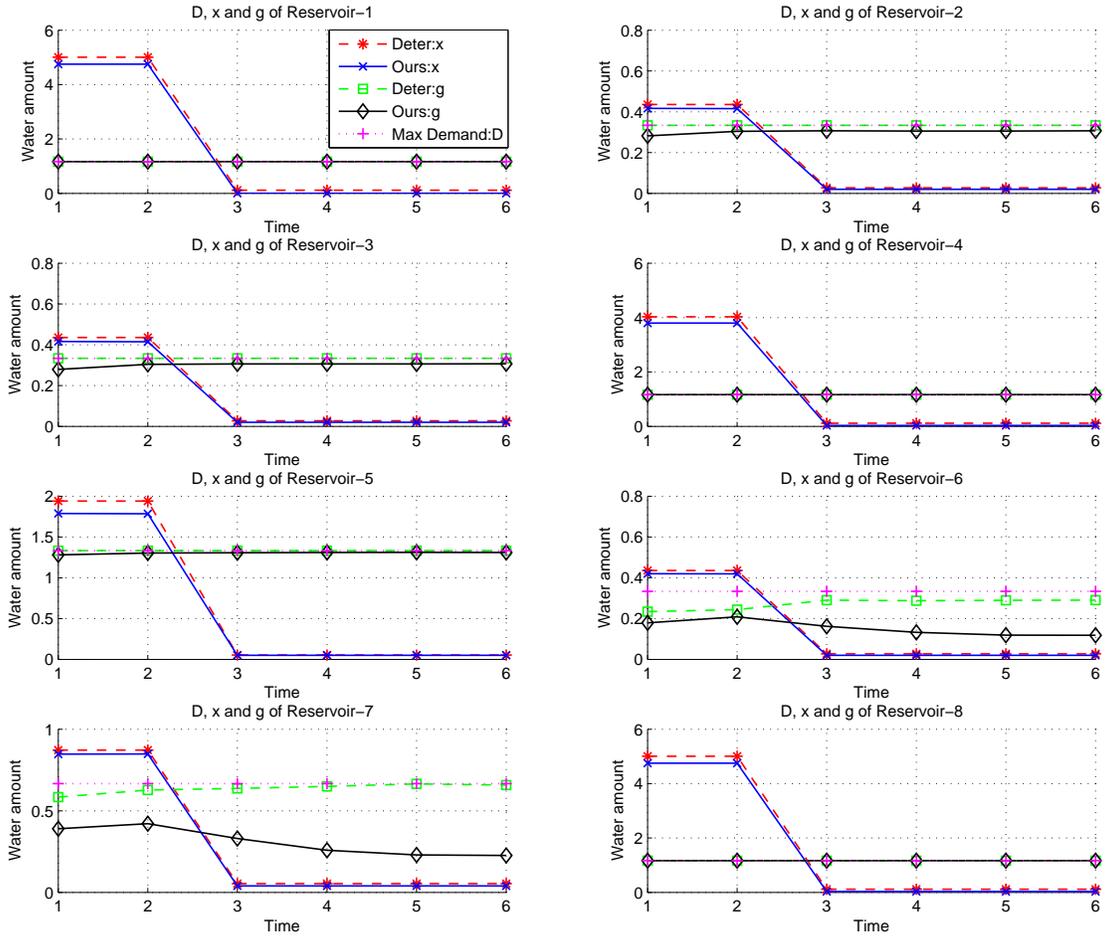}} \vskip -0.35in
\caption{Inflow prediction $x_n^t$, released volume $g_n^t$, and
$D^t_n$, for each method and for each reservoir.}
\label{fig_complex2}
\end{center}
 \vskip -0.35in
\end{figure}

Currently,  by royal initiating of his Majesty the King of Thailand,
a multi-connection reservoir network, namely Ang-Puang, has been
being constructed in Petchburi province,
Thailand\footnote{http://www.huaysaicenter.org/water.php}. The
geography of the network is shown in Figure~\ref{fig_angpuang}, and
the network topology, together with the capacity of each reservoir
(in metric tonne), is shown in Figure~\ref{fig_reservoir_exp}
(Right). For simplicity, we will mention each reservoir by its order
number, e.g. reservoir-1, reservoir-2, etc., instead of its name.
Since the network is currently being constructed and all the
historical data are being collected, in this experiment we are able
to employ only artificial data. The artificial data employed here is
in the same spirit as the standard 4 and 10 reservoir problems for a
traditional network
\cite{Larson:Book68,Yakowitz:WRR79,Wardlaw:JWRPM99,Janejira:Paddy05,Jalali:WRM07}.
Nevertheless, our artificial data contains also stochastic
information while there is none in the 4 and 10 reservoir problems.
Here, $C_n^t(\cdot)$, $R_n^t(\cdot)$ and $G_n^t(\cdot)$ are the same
as the previous case, illustrated in Figure~\ref{fig_CRG}. The
following values are specified: $T = 6$, $N =8$, $Q_{nm} = 2.5$,
$G_n^t = 1$, $C_{nm}^t = 0.25$ and $R_n^t = 2.5$  for all $n,m,t$.
Other parameters are shown in Figure~\ref{fig_complex1} and
Figure~\ref{fig_complex2}, i.e. Figure~\ref{fig_complex1}
illustrates $Pr(x^t_n)$ for $t=1,2$ and $v^0_1 = v^0_2 = V_1 = V_2$,
and Figure~\ref{fig_complex2} shows the values of ``maximum demand''
$D$ specified in $G_n^t(\cdot)$ for each reservoir in each period.

This problem setting is a simplified version of a real situation
normally occured in Thailand where $t$ is a two-month period and
$1,...,T$ is thus one year. A long-term plan is determined just
before the rainy season; thus, only for $t=1,2$ where $x^t_n$ will
have probabilities to be significantly more than zero, and almost
zero in the $t=3,...,6$ periods. To save space, $Pr(x^t_n)$ for
$t=3,...,6$ (which their realizable values are around zeros) are not
shown. The intuition of this problem setting is that, at the first
two periods $t=1,2$, because of the huge amount of rain, all
reservoirs will, with high probability, nearly attain their maximum
capacities. However, only the big reservoirs $n=1,4,8$ will have
enough water for later periods and are responsible to transfer water
to the other small reservoirs.

The total profit of the two methods are shown in
Figure~\ref{fig_complex1} (Down-Right). Note that, for both methods,
standard deviations are almost zeros, and thus our method
significantly provides more profit than the conventional
deterministic method. As can be seen from Figure~\ref{fig_complex2},
in this situation, our method usually becomes more conservative
about amount of inflow than the conventional method. As a result,
our method cannot guarantee water supply in reservoir-6 and
reservoir-7; in contrast, the plan of conventional LP decide to
provide almost enough water for every reservoir in every period.
Nevertheless, there are often cases that the real inflow
$\tilde{x}^t_n$ for each reservoir is not as high as expected from
$x^t_n$ of the plan of conventional LP; thus, a high risk cost has
often to be paid for extra water requirement.

It is also important to note that, in the simulations, the
time-complexity of our method increases only slightly from
traditional LP. By using \textsc{matlab}'s ``linprog'' function with
the ``largescale'' option. Our method converges in 18 iterations
(4.52 seconds) where traditional LP converges in 17 iterations (4.28
seconds).

\subsection{Sensitivity Analysis}
\begin{figure*}[t]  \begin{center} \vskip -0.5in \hbox{\hskip -0.2in
\setlength{\epsfxsize}{3.1in} \epsfbox{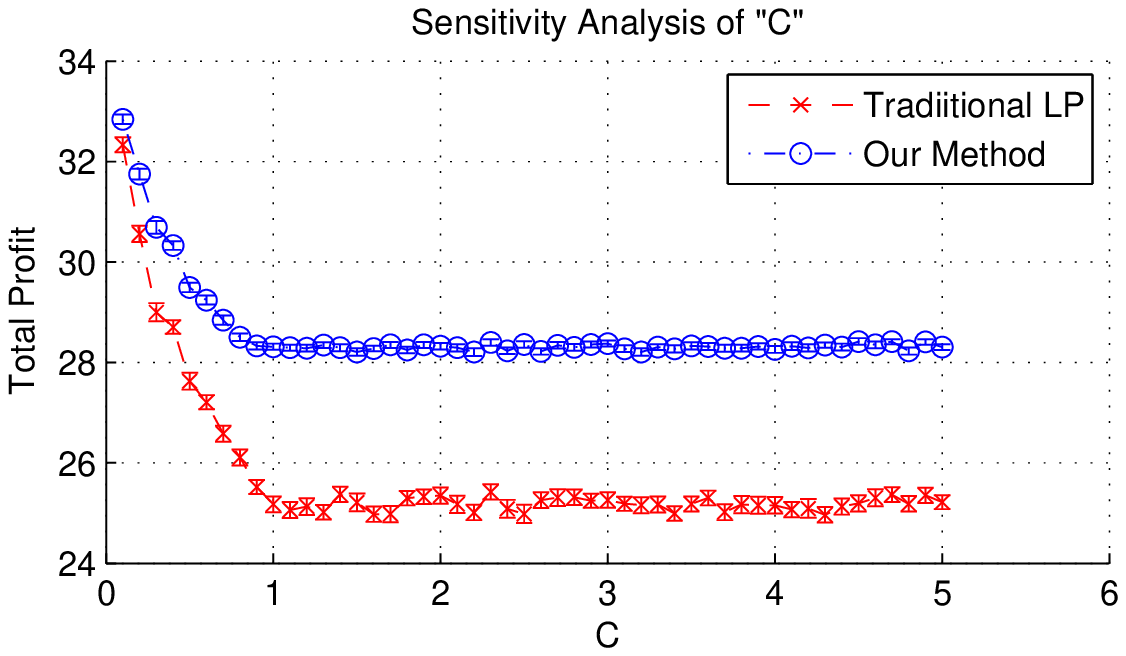}
\setlength{\epsfxsize}{3.1in} \epsfbox{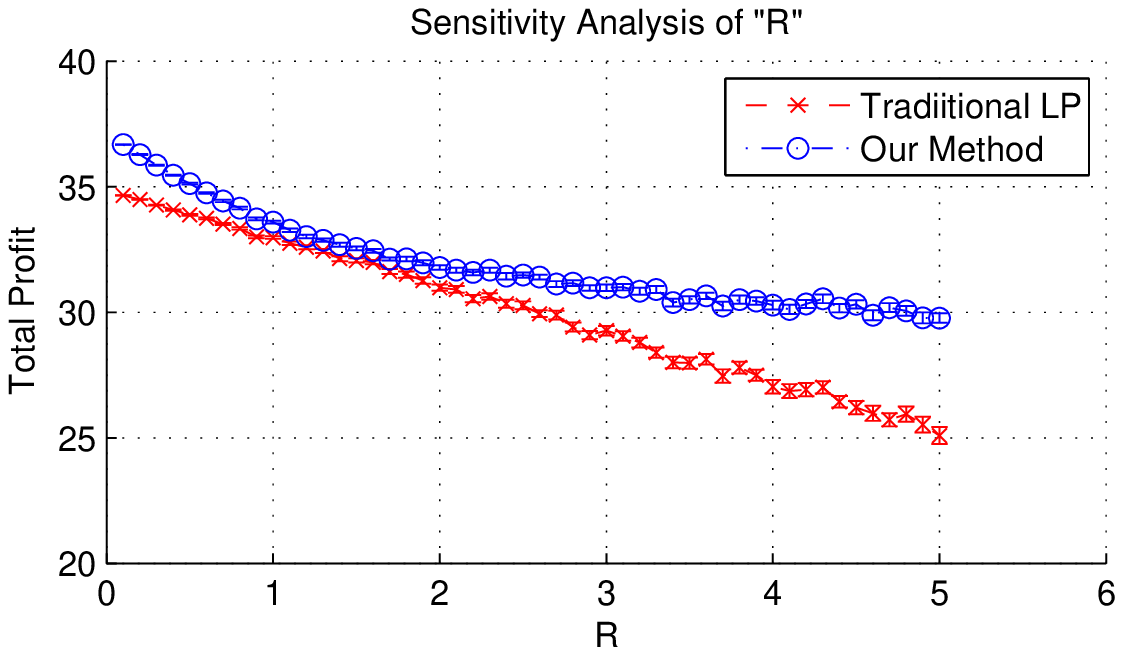} }
\end{center}
\begin{center} \vskip -0.2in \hbox{\hskip -0.2in
\setlength{\epsfxsize}{3.1in} \epsfbox{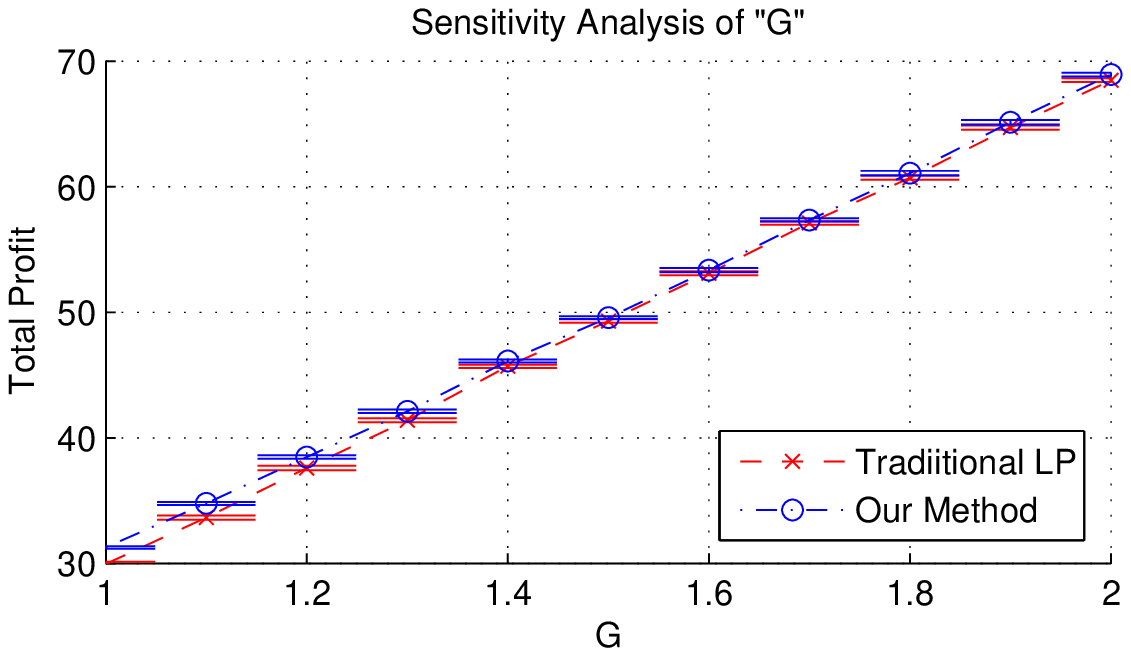}
\setlength{\epsfxsize}{3.1in} \epsfbox{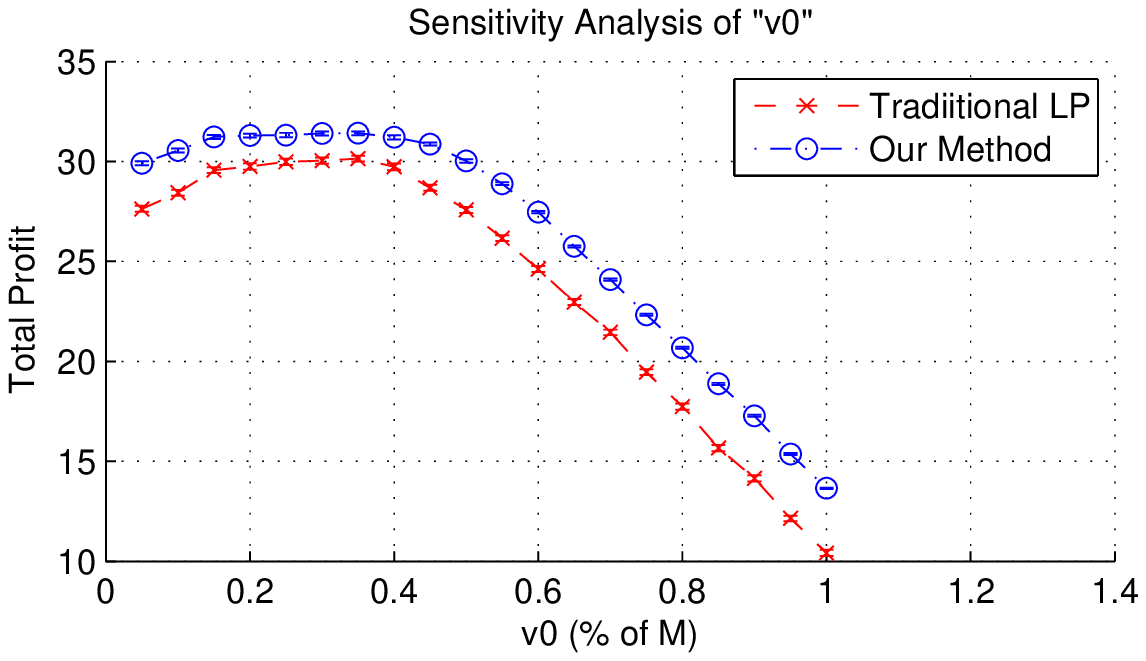} }
\end{center}
\vskip -0.5in \caption{Sensitivity analysis of the two methods.}
\label{fig_sensitive} \vskip -0.0in
\end{figure*}

In this section, we demonstrate the sensitivity of each method with
respect to four parameters of the complex-network problem explained
in Section~\ref{sect_exp2}. Figure~\ref{fig_sensitive} shows results
of the sensitivity analyzes. In each analysis, all the parameters
are the same as those of the experiment in Section~\ref{sect_exp2},
except the analyzed parameter written in the title of each figure.
Each figure shows the average total profit of each method in each
parameter setting. The standard deviation of the average is also
drawn; nevertheless, the standard deviation of each point is very
small and is difficult to observe in every figure. This indicates
that the statistical confidences of the results shown here are very
high. Note that our method always significantly outperforms the
existing method in all cases.

Figure~\ref{fig_sensitive} (Up-Left) illustrates the case where the
parameter `$C_{nm}^t$' is varied. We can see that in the beginning
where $C_{nm}^t \le 1$, the total profits of the two methods
decrease about linearly with respect to the increase of $C_{nm}^t$.
After $C_{nm}^t \ge 1$, the total profits do not change further
since the transfer cost becomes more than the release profit
$G_{nm}^t$, and thus no water transfer occurs after $C_{nm}^t > 1$.

Figure~\ref{fig_sensitive} (Up-Right) illustrates the case where the
parameter `$R_{n}^t$' is varied. We can see that the average total
profit of traditional LP decreases about linearly with respect to
the increase of $R_{n}^t$. In contrast, the average total profit of
our method decreases in a slower rate. This is because our method is
able to efficiently take the risk and the stochastic information
into account while traditional LP cannot. As a result, our method is
less sensitive to the change of $R_{n}^t$ than traditional LP.

Figure~\ref{fig_sensitive} (Down-Left) illustrates the case where
the parameter `$G_{n}^t$' is varied. In this case, the average total
profits of the two methods  increase about linearly with respect to
$G_{n}^t$. In this case, like other cases, our method provides
significantly higher total profit than traditional LP although it is
difficult to see in the figure due to the large scale of the y-axis.

Figure~\ref{fig_sensitive} (Down-Right) illustrates the case where
the parameter `$v_{n}^0 = V_n$' is varied for each reservoir. In
this case, the $v_{n}^0$ is varied from 5\% of $M_n$ to 100\% of
$M_n$. As a result, the sensitivity of each method is about the
same. Note that as the value of $v_{n}^0  = V_n$ becomes near $M_n$,
the total profits decrease since the constraint $v_{n}^T \ge V_n$
forces the optimizer not to use any water at later periods.


\section{Discussions}
As we explained earlier, a deterministic method seems to be the most
appropriate among previously existing methods for the problem of
demand-supply optimization in a multi-connection network considered
in this paper. From the experiments, we have shown that our new
optimization framework gives promising results by always
outperforming the existing deterministic method in all experiment
settings.

There are some issues which should be considered in practical uses
of our method. Firstly, remember that our method requires to
discretize $Pr(x_n^t)$. This discretization is not an important
limitation of our method since, in practice, all statistics from
historical data are in fact originally discrete. Hence, the original
statistical distributions can be applied to our method directly.
Nevertheless, an application of a smoothing method might be useful
to obtain simplified probability distributions which can reduce an
optimization time of our method.

Secondly, we note that although the class of convex programming is
efficiently solvable, the class of linear programming is still the
most preference in term of time efficiency. In fact, theoretically,
a convex function can be linearized which an arbitrary degree of
approximation accuracy. Therefore, in our opinion, one potential
future research direction is to apply an efficient linearization
method from the computer vision community to linearize a convex
objective function \cite{Salotti:PRL01,Kolesnikov:PRL03}, and find
the best tradeoff between the time efficiency and the approximation
accuracy.

Finally, it is important to note that although our method provides a
static long-term plan, our method can be used to dynamically update
a plan as well (by re-running the method in the beginning of each
time period), provided that the length of $t$ is not shorter than a
running time of our method. In all experiments mentioned in
Section~\ref{sect_exp}, the optimization process of our method
usually accomplished within 1 minute; therefore, for any long-period
plans, e.g. $t$ is a week or a month, our method can also provide a
dynamic plan.





\subsubsection*{Acknowledgements}
This work was accomplished while the first author stayed at Kanda
Laboratory, Ookayama Campus, Tokyo Institute of Technology, and was
supported by Japan Student Services Organization (JASSO) under the
JENESYS program. We are grateful to Prof. Manabu Kanda and Prof.
Kumiko Yokoi for hosting the first author while he was doing this
research. We also thank Makoto Nakayoshi and Dr. Kanlaya
Sunthornwongsakul for helpful discussions.

\bibliography{reservoir}
\bibliographystyle{theapa}
\end{document}